\documentclass[reprint,amsmath,amssymb,aps]{revtex4-2}

\usepackage{graphicx}
\usepackage{dcolumn}
\usepackage{bm}
\usepackage{amsmath}
\usepackage[usenames,dvipsnames]{xcolor} 
\usepackage{mathtools}
\usepackage[colorlinks=true,citecolor=blue,linkcolor=blue]{hyperref}
\usepackage[caption=false]{subfig}
\usepackage{bbm}
\usepackage{braket}
\usepackage{physics}
\usepackage{comment}
\usepackage{hyperref}
\usepackage{upgreek}
\usepackage{esint}
\usepackage{soul}
\usepackage[normalem]{ ulem } 
\usepackage{floatrow}

\newcommand{\leri}[1]{\left(#1\right)}

\begin{document}

\title{Programmable quantum simulations on a trapped-ions quantum simulator with a global drive}

\author{Yotam Shapira$^{1,\ast}$}
\author{Jovan Markov$^{1,\ast}$}
\author{Nitzan Akerman$^{1}$}
\author{Ady Stern$^{2}$}
\author{Roee Ozeri$^{1}$}

\affiliation{\small{$^1$Department of Physics of Complex Systems, Weizmann Institute of Science, Rehovot 7610001, Israel\\
$^2$ Department of Physics of Condensed Matter Systems, Weizmann Institute of Science, Rehovot 7610001, Israel\\
$^*$ These authors contributed equally to this work
}}

\begin{abstract}
    Simulation of quantum systems is notoriously challenging for classical computers, while quantum hardware is naturally well-suited for this task. However, the imperfections of contemporary quantum systems poses a considerable challenge in carrying out accurate simulations over long evolution times. Here we experimentally demonstrate a method for quantum simulations on a small-scale trapped ions-based quantum simulator. Our method enables quantum simulations of programmable spin-Hamiltonians, using only simple global fields, driving all qubits homogeneously and simultaneously. We measure the evolution of a quantum Ising ring and accurately reconstruct the Hamiltonian parameters, showcasing an accurate and high-fidelity simulation. Our method enables a significant reduction in the required control and depth of quantum simulations, thus generating longer evolution times with higher accuracy.
\end{abstract}

\maketitle

Quantum simulators are controllable quantum systems that enable the study of phases, dynamics, and properties of complex quantum systems, for which an analytical or numerical treatment is challenging \cite{feynman1982simulating}. Quantum simulations are considered a suitable task for noisy intermediate scale quantum (NISQ) computers \cite{preskill2018quantum}, that are currently available. Recent years have seen numerous implementations of quantum simulations on several  platforms \cite{neill2021accurately,kim2023evidence,semeghini2021probing}, which are performed with an ever-increasing quality, approaching an advantage over classical methods \cite{kechedzhi2023effective}. An apparent challenge to NISQ-era quantum hardware is to perform large-scale quantum simulations, with a relatively shallow circuit depth, i.e. with few operations, in order to avoid the deterioration of the simulation's fidelity due to noise.

Ion-crystals trapped in radio frequency (RF) traps are an especially prolific tool for quantum simulations \cite{manovitz2020quantum,shapira2023quantum,kokail2019self,joshi2022observing,tan2021domain,kyprianidis2021observation,monroe2021programmable,qiao2022observing,iqbal2023creation,porras2004effective,wu2023qubits,lu2023realization,hayes2014programmable,katz2024observing}, due to their long coherence times \cite{wang2017single,wang2021single}, high-fidelity control \cite{gaebler2016high,clark2021high,ballance2016high,srinivas2021high} and rich connectivity \cite{shapira2020theory,shapira2023fast,grzesiak2020efficient,wang2022fast}. By exploiting the long-range coupling between all ions in the ion-crystal, it is possible to perform parallel and long-range entanglement, potentially increasing the efficacy of the simulation. Analog quantum simulation in trapped-ions platforms are typically either limited to simple models that are constrained by the one-dimensional (1D) structure of the underlying ion-crystal \cite{porras2004effective}, or require  local control \cite{hayes2014programmable}.   

Here we experimentally implement quantum simulations of the Ising-spin model on a small-scale trapped ions quantum system \cite{manovitz2022trapped}. We evolve our system using global pulses, that drive the ions homogeneously, and nevertheless generate a desired inhomogeneous programmable interaction in each pulse, which is unconstrained by the 1D linear geometry of the ion-chain. This is enabled by coherently and simultaneously coupling to all modes of motion of the trapped ions crystal, in a controllable manner. We are able to simulate spin-Hamiltonians of the form,
\begin{equation}
    H=\sum_{n,m=1}^N J_{n,m}\sigma_x^{\leri{n}}\sigma_x^{\leri{m}},\label{eqHMulti}
\end{equation}
with $\sigma_x^{\leri{n}}$ a Pauli-$x$ operator acting on the $n$th spin of a $N$ site spin system, and $J_{n,m}$ an experimentally controllable coupling matrix. This implements an Ising-spin model.

With global methods that are straightforward in trapped ions systems, we extend this interaction to accommodate for a transverse field, $\delta\sum_{n=1}^N\sigma_z^{\leri{n}}$. Similarly, we can add  $\sigma_y^{\leri{n}}\sigma_y^{\leri{m}}$ terms, that may differ in their coupling matrix, compared to the coupling in Eq \eqref{eqHMulti}. 

Our method takes into account and mitigates unwanted inhomogeneous aberrations due to, e.g. finite driving beam waist. Furthermore, while all modes of motion of the ion-crystal may be used, here we explicitly decouple from the center-of-mass mode as it is more prone to heating and decoherence, without affecting our method's programmability, thus improving the simulation's fidelity.

We simulate a 4-ions Ising-spin quantum ring and observe dynamics under its Hamiltonian. We supplement this model with a transverse field, such that the resulting evolution is purely quantum mechanical, and observe the dynamics as a function of the transverse field magnitude. 

Below we analyze these results in full, and are able to accurately reconstruct both the Ising coupling terms, $J_{n,m}$, as well as the various transverse fields, $\delta$, showcasing an accurate realization of quantum simulations using our method.

The theoretical proposal underpinning our work has been proposed in Refs. \cite{shapira2020theory,shapira2023fast}. We provide the relevant physical picture and method details. In trapped-ions-based quantum systems, entanglement between qubits is typically generated by spin-dependent forces, which mediate spin-spin interactions through the collective phonon modes of motion of the ion-crystal. Conventionally only two ions are driven and coupled to a single phonon mode, such as in the M\o lmer-S\o rensen (MS) gate \cite{sorensen1999quantum,sorensen2000entanglement}, generating an evolution of the form $U_\text{MS}=\exp\leri{i\Phi\sigma_x^{\leri{n}}\sigma_x^{\leri{m}}}$, with $n$ and $m$ the indices of the two entangled ions and $\Phi$ a controllable entanglement phase. 

This method is generalized by homogeneously driving all of the $N$ ions in the ion-crystal. We purposefully couple to $N$ modes of motion along the axial axis of the ion-crystal. As shown in Ref. \cite{shapira2020theory}, at a fixed pulse duration, such a drive yields the evolution,
\begin{equation}
    U=\exp\leri{i\sum_{j=1}^N\Phi_j\sum_{n,m=1}^N O_j^{\leri{n}}O_j^{\leri{m}}\sigma_x^{\leri{n}}\sigma_x^{\leri{m}}},\label{eqUMulti}
\end{equation}
with the $\left\{\Phi_j\right\}_{j=1}^N$ completely controllable mode-dependent entanglement phase, and $O$ an orthonormal mode-matrix, i.e $O_j^{\leri{n}}$ is the normalized participation of the $n$th ion in the $j$th mode of motion \cite{james1998quantum}.

We drive the ions in the adiabatic regime, i.e. the pulse duration implementing $U$ is slower than the characteristic oscillation periods of the modes of motions. In this regime the mapping between the the set of desired phases, $\left\{\Phi_j\right\}_{j=1}^N$, and the details of the drive is straightforward, as described below. Faster realizations of $U$ are discussed in Ref.\cite{shapira2023fast}.

Our drive is composed of 4 tones per motional mode: $\omega_{j,1,\pm}=\omega_0\pm\leri{\nu_j+\xi}$ and $\omega_{j,3,\pm}=\omega_0\pm\leri{\nu_j+3\xi}$, with $\nu_j$ the frequency of the $j$th mode of motion and $\omega_0$ the single-qubit transition frequency. $\xi$ is an additional detuning from the mode transition frequencies such that  $U$ in Eq. \eqref{eqUMulti} is achieved at $t=2\pi/\xi$. 

The phases of the tones of $\omega_{j,1,\pm}$ are set to 0 (for all $j$'s) and the phases of the tones $\omega_{j,3,\pm}$ are set to $\pi$. With these choices the resulting evolution is made robust to unwanted coupling to the carrier transition and pulse-timing errors \cite{shapira2018robust,shapira2020theory,markov2023digital}.

The amplitudes of $\omega_{j,n,\pm}$, driving the $j$th mode of motion are set to $r_j$. With these generalizations of the MS gate we make use of consecutive implementations of $U$ in Eq. \eqref{eqUMulti}, such that the ion's evolution, at integers multiples of $2\pi/\xi$, evolves stroboscopically according to the Hamiltonian in Eq. \eqref{eqHMulti}, and $J_{n,m}\propto\sum_{j=1}^N \frac{\eta_j^2 r_j^2}{\xi} O_j^{\leri{n}}O_j^{\leri{m}}$, with $\eta_j$ the Lamb-Dicke parameter of the $j$th mode of motion. The proportionality constant accommodates for the laser's total Rabi frequency, $\Omega_0$, and additional (mode and ion independent) numerical factors.

Due to the orthonormality of $O$, shifting all $\Phi_j$s by a constant, i.e. $\Phi_j\mapsto\Phi_j+\phi$, amounts to modifying $U$ in Eq. \eqref{eqUMulti} by a global phase, i.e. $U\mapsto e^{i N \phi}U$ \cite{SM}. Here we exploit this property in order to set $\Phi_{j=1}=0$, and adjust all the other $\Phi_j$s appropriately. This is helpful as the $j=1$ mode is an axial center-of-mass mode, which typically exhibits the fastest heating and decoherence rate. Thus, we decouple from this mode, yielding longer coherence times and higher simulation fidelity.

Furthermore, inhomogeneities in the field driving the ions, e.g. due to a finite beam waist of an optical drive, can be taken into account by adjusting $O$ accordingly \cite{SM}, enabling  mitigation of such effects.

The Pauli-$x$ rotations, i.e. $\sigma_x^{\leri{n}}$,  in Eq. \eqref{eqUMulti} can be trivially generalized to $\sigma_\phi^{\leri{n}}=\cos\leri{\phi}\sigma_x^{\leri{n}}+\sin\leri{\phi}\sigma_y^{\leri{n}}$, with $\phi$ fully controllable, by equally shifting the phases of the drive's tones. The choice of $\phi$, as well as all other drive parameters such as the different $\Phi_j$'s, can be changed at each consecutive implementation of $U$. Combined with the well-known Suzuki-Trotter decomposition \cite{trotter1959product,suzuki1976generalized}, this accommodates for various spin-Hamiltonians, e.g.
\begin{equation}
    H=\sum_{n,m=1}^N\left[J_{n,m}^{\leri{x}}\sigma_x^{\leri{n}}\sigma_x^{\leri{m}}+J_{n,m}^{\leri{y}}\sigma_y^{\leri{n}}\sigma_y^{\leri{m}}\right]+\delta\sum_{n=1}^N\sigma_z^{\leri{n}},\label{eqHGeneral}
\end{equation}
with $J_{n,m}^{\leri{x}}$ and $J_{n,m}^{\leri{y}}$ controllable couplings and $\delta$ a controllable transverse field. The $\sigma_y^{\leri{n}}$ terms are generated by a $\pi/2$ jump in $\phi$, while the transverse $\sigma_z^{\leri{n}}$ terms are generated by a gradual linear ramp of $\phi$ in each consecutive Suzuki-Trotter block. 

The general Hamiltonian in Eq. \eqref{eqHGeneral}, as well as more elaborate time-dependent Hamiltonians, are all made possible while still using a global driving field. We remark that $H$ commutes with the parity operator $P=\exp\leri{i\pi\sum_{n=1}^N\sigma_z^{\leri{n}}}$, thus states with well-defined parity, e.g. superpositions of states with an even number of spin-excitations, will remain with the same parity throughout their evolution. 

We experimentally implement quantum simulations on a small-scale trapped-ions quantum simulator \cite{manovitz2022trapped}, in which the $\ket{5S_{\frac{1}{2},\frac{1}{2}}}$ ($\ket{4D_{\frac{5}{2},\frac{3}{2}}}$) states of $^{88}\text{Sr}^+$ ions are mapped to $\ket{0}$ ($\ket{1}$) qubit levels. These levels are coupled, using a quadrupole transition, by a 674 nm narrow linewidth laser \cite{peleg2019phase}, illuminating the ions approximately homogeneously with a wide global beam. We modulate the 674 nm beam appropriately such that it has a rich spectrum, containing $M$ frequency pairs, $\left\{\omega_0\pm\omega_m\right\}_{m=1}^M$, with appropriate amplitudes and phases, as prescribed above.

We first investigate dynamics under a 4-site spin ring, with anti-periodic boundary conditions, i.e. the nearest-neighbor (\textit{n.n}) Hamiltonian,
\begin{align}
\begin{split}
H_\text{ring,a.p}&=\Omega\leri{\sigma_x^{\leri{1}}\sigma_x^{\leri{2}}+\sigma_x^{\leri{2}}\sigma_x^{\leri{3}}+\sigma_x^{\leri{3}}\sigma_x^{\leri{4}}-\sigma_x^{\leri{4}}\sigma_x^{\leri{1}}} \\ & +\delta\sum_{n=1}^4\sigma_z^{\leri{n}},\label{eqHRingApbc}
\end{split}
\end{align}
with `anti-periodicity' manifested as the negative sign of the $\sigma_x^{\leri{4}}\sigma_x^{\leri{1}}$ coupling term. 

This Hamiltonian is analytically solvable, thus we use it as a proof of concept for highlighting key components of our method. Namely freely generating positive and negative coupling terms, long-range terms that are not restricted by the 1D underlying ion-crystal and correction of aberrations due to finite beam width. Furthermore it is physically motivated, since with a Jordan-Wigner \cite{shankar2017exact} transformation the system is mapped to an even number of fermions hopping on a periodic ring (with no negative couplings) \cite{butiker1983josephson,cheung1988persistent}.

$H_\text{ring,a.p}$ can be well-approximated by the axial modes of motion of our harmonic ion-trap. The method for choosing the appropriate entanglement phases is detailed in \cite{shapira2020theory,SM}, yielding, $\boldsymbol{\Phi}^{\leri{\text{optimal}}}=\leri{0,0.3867,-0.7071,-1.0939}$, with an ideal implementation fidelity of $H_\text{ring,a.p}$ of 0.989 (defined precisely below). We note that Ref. \cite{kyprianidis2024interaction} provides a systematic analysis of the possible couplings schemes enabled by global modes and that Refs. \cite{li2022realizing,espinoza2021engineering} further expand these possibilities by directly shaping the global mode structure.

\begin{figure}
    \centering
    \includegraphics[width=1\linewidth]{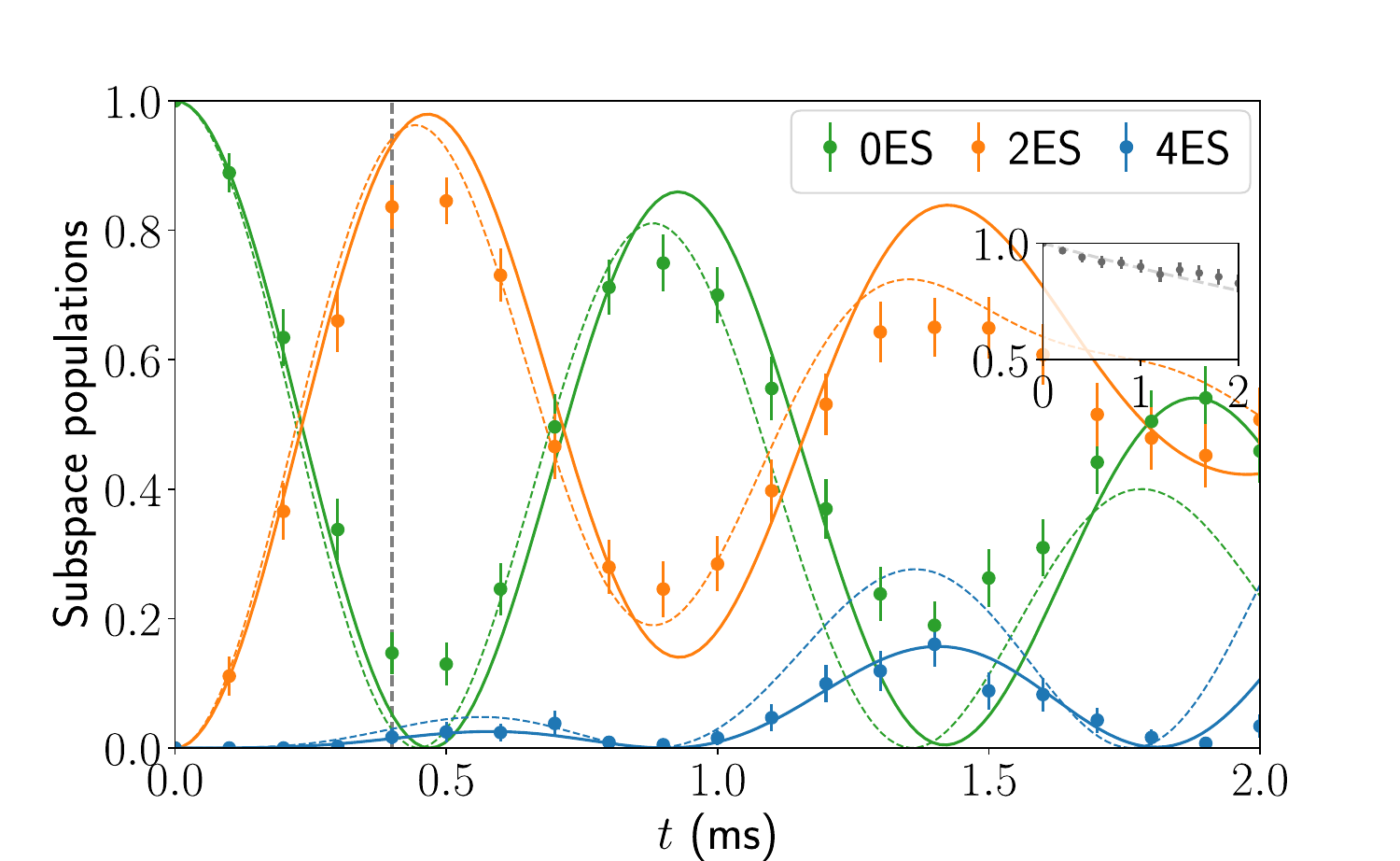}
    \caption{Excitation subspace dynamics of the anti-periodic Ising-spin ring. Populations are measured at consecutive implementations of $U$ (points), and grouped into excitation subspaces (colors). The theoretical prediction (dashed) shows a good agreement with the data. A more involved model that includes dephasing (solid) is fitted to the data yielding a mean fit error of $<0.155$. Populations in the even excitation subspaces (green, orange and blue) exhibit coherent oscillations. The inset shows the total (unnormalized) population in the even subspaces (points), which slowly decreases due to decoherence and is well fitted by our full model (solid)}. Error bars reflect $\pm2\sigma$ statistical errors due to quantum shot-noise.
    \label{figDynamics}
\end{figure}

We first benchmark our simulation with $\delta=0$. To do so we initialize the system to the ground state, $\ket{0000}$, evolve it under $H_\text{ring,a.p}$, and measure the population at the different spin states. Figure \ref{figDynamics} shows the state occupations after such an evolution, with the spin populations grouped in terms of their respective excitation subspaces (ES), i.e. the number of excitations in each state. The evolution is sampled stroboscopically, at integer multiples of $U$, yielding data (points) with error bars that reflect $\pm2\sigma$ statistical errors due to quantum projection noise. Here and in the measurements below we use $\xi=5$ kHz. Since the initial state is the ground state we expect the evolution to remain in even subspaces 0ES (green), 2ES (orange) and 4ES (blue). Thus we post-select and normalize to the population in even subspaces.

The data is shown together with the ideal theoretical prediction of the evolution under $H_\text{ring,a.p}$ (dashed lines). As previously analyzed in Ref. \cite{shapira2023quantum}, discrepancies between the data and theoretical prediction, are attributed to qubit dephasing ($T_2\approx5$ ms). We account for it, as well as inaccuracies of the $J_{n,m}$'s, using a Lindblad master equation, and use it to fit the six independent parameters of $J_{n,m}$ and $T_2$.  The fit matches the populations measured over all 8 even-excitation states, with an average population error of $<0.0155$ per state, per data point, which is consistent with shot-noise errors under 500 experimental repetitions per-point, and a characteristic coupling rate of $\tilde{\Omega}\approx1.13$ kHz (further details in \cite{SM}). 

Furthermore, this analysis predicts population 'leakage' from the even excitation subspaces (inset of Fig. \ref{figDynamics}), as well as deviations between the ideal and measured evolution, all of which are accounted for by qubit dephasing. Thus improving $T_2$ will enable longer simulations with greater fidelity.

Various methods exist to validate the performance of our simulation \cite{granade2012robust,bairey2019learning,benav2020direct}. Here we directly benchmark our simulation by performing parity measurements. Such a method has been recently employed in Ref. \cite{guo2024hamiltonian}. We do so by evolving the system for a short time, $t_\text{cor}=400\mu\text{s}$ (vertical dashed gray in Fig. \ref{figDynamics}), corresponding to two consecutive implementations of $U$, and then applying a global `analysis' $\pi/2$-pulse with phase $\phi$. That is, the system evolves according to,
\begin{equation}
    U_\text{cor}=e^{\frac{i}{2}\frac{\pi}{2}\sum_{n=1}^N\sigma_\phi^{\leri{n}}} e^{i\Omega t_\text{cor}\sum_{n,m}J_{n,m}\sigma_x^{\leri{n}}\sigma_x^{\leri{m}}}\label{eqUParity},
\end{equation}
after which we measure state occupations and evaluate the bipartite-correlations, $\langle\sigma_z^{\leri{n}}\sigma_z^{\leri{m}}\rangle$. This process constitutes a measurement of $C_{n,m}\leri{\phi}=\Braket{\psi\leri{t}|\sigma_{\phi^\prime}^{\leri{n}}\sigma_{\phi^\prime}^{\leri{m}}|\psi{\leri{t}}}$, with $\phi^\prime=\phi-\pi/2$ and $|\psi\leri{t}\rangle=U\leri{t}|0000\rangle$. For short evolution times, $\Omega t_\text{cor}J_{n,m}\ll1$, we get $C_{n,m}=-2\Omega t_\text{cor}J_{n,m}\sin\leri{2\phi}$ with corrections that are cubic in $\Omega t_\text{cor}$ \cite{SM}.

\begin{figure}
    \centering
    \includegraphics[width=0.9\linewidth]{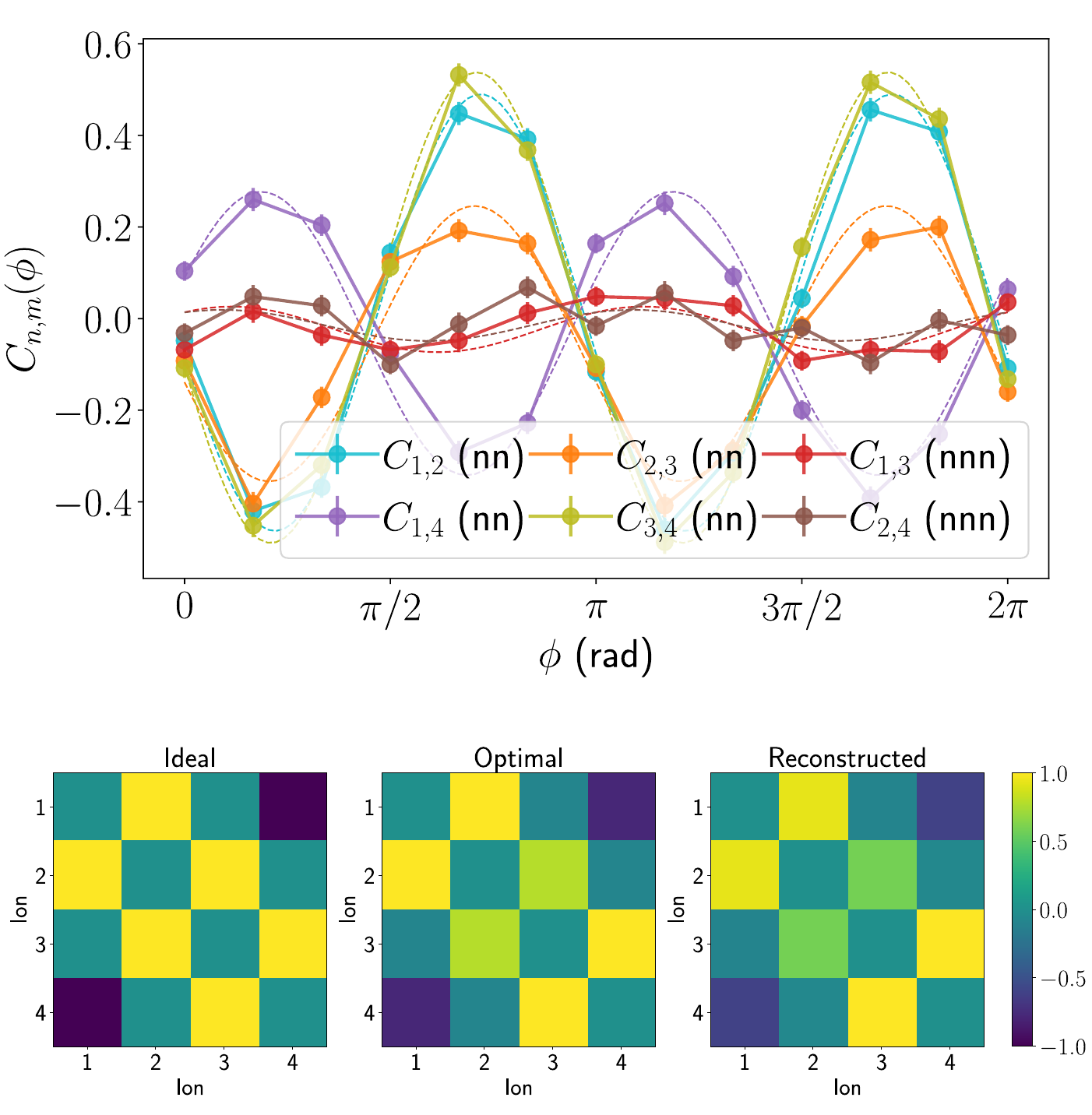}
    \caption{Top: The correlation $C_{n,m}=\langle{\sigma_\phi^{\leri{n}}\sigma_\phi^{\leri{m}}}\rangle$ is evaluated for pair of sites, at various values of $\phi$, after a small evolution time $t_\text{parity}=6\pi\xi^{-1}$. We observe high-contrast oscillations between sites that are \textit{n.n}. The anti-periodic coupling of $\sigma_x^{\leri{1}}\sigma_x^{\leri{4}}$ in Eq. \eqref{eqHRingApbc}, is manifested as an opposite-phase fringe of $C_{1,4}$ (purple). Error bars reflect $\pm2\sigma$ statistical errors due to quantum projection-noise. Data (points) is fitted to a single oscillating sine (dashed) showing a good fit. Bottom: coupling matrices, $J_{n,m}$, of the anti-periodic Ising-spin ring. Showing ideal values (left) read-off the model's Hamiltonian in Eq. \eqref{eqHRingApbc}, optimal values (middle) due to a global drive implementation, and reconstructed values (right), using the parity fits (top). The matrices are normalized such that the largest entry in each of them is $1$ (arbitrary units). }
    \label{figParity}
\end{figure}

The results of these measurements, for various rotation phases $\phi$, are shown in Fig. \ref{figParity} (top). We observe that correlations that are \textit{n.n.} on the spin-ring (cyan, orange, olive, and purple) exhibit high-contrast fringes, while next nearest-neighbour (\textit{n.n.n.}) sites (red and brown) do not. Due to the anti-periodic boundary conditions, $C_{1,4}$ (purple) exhibits an oscillation phase which is opposite to all other pairs.
Variations of the global drive on the ions may reduce the fidelity of the analysis pulse and bias our data. However this effect is quadratic in the field inhomogeneity \cite{SM}, and may be further mitigated with well-known amplitude-robust composite pulses.  Here we simply rescale the fringe contrasts, using the relative local Rabi frequencies in the ion-crystal, constituting a  $<1\%$ effect to the data.

We fit each correlation fringe to a sine function with frequency $2\phi$ (dashed), yielding a mean fitting error of 0.018, well within error bars. The resulting fitted amplitudes,  $J_{n,m}^\text{reconstructed}$, are presented in Fig. \ref{figParity} (bottom, right), along with the ideal values of $J_{n,m}^{\text{ideal}}$ (bottom, left) which are directly read-off the model's Hamiltonian in Eq. \eqref{eqHRingApbc}, as well as the expected values of the implemented model, $J_{n,m}^{\text{optimal}}$ (bottom, middle), originating from our choice $\boldsymbol{\Phi}^{\leri{\text{optimal}}}$, and the system's modes of motion \cite{james1998quantum,SM} (middle). 

To quantify our reconstruction, we consider the Cartesian overlap of the non-trivial values of the various $J$s. Specifically we define $F\leri{J_1,J_2}\equiv\boldsymbol{J}_1\cdot\boldsymbol{J}_2/\sqrt{\leri{\boldsymbol{J}_1\cdot\boldsymbol{J}_1}\leri{\boldsymbol{J}_2\cdot\boldsymbol{J}_2}}$, with $\boldsymbol{J}_1,\boldsymbol{J}_2$ being 6-element real vectors constructed from the upper triangular entries of the coupling matrices $J_1$ and $J_2$ (excluding the trivial diagonal). This metric corresponds to the Hilbert-Schmidt fidelity between Hamiltonians, $F_\text{HS}\leri{H_1,H_2}=\left|\text{Tr}\leri{H_1 H_2}\right|/\sqrt{\text{Tr}\leri{H_1^2}\text{Tr}\leri{H_2^2}}$, when considering Hamiltonians constructed by $\sigma_x^{\leri{n}}\sigma_x^{\leri{m}}$-type interactions, which is natural in our setup. The denominator makes the comparison relative, i.e. up to a rate factor, $\Omega$.  

The three reconstructed coupling matrices are in good agreement. Indeed the overlap yields $F\leri{J^\text{ideal},J^{\text{optimal}}}=0.989$ and $F\leri{J^{\text{optimal}},J^\text{reconstructed}}=0.993\substack{+0.007 \\ -0.042}$, indeed verifying a high-quality implementation of the intended ring model. These high-fidelity results may seem contradictory to measured dynamics in Fig. \ref{figDynamics}, where the ideal (dashed) values deviate significantly from the data (points) or fit (solid). However, this is simply a manifestation of the well-known robustness of correlation measurements to uncorrelated noise \cite{shaniv2018toward,shaniv2019quadrupole,manovitz2019precision}.

We now turn to study the effects of the transverse field, $\delta$, in the ring Hamiltonian in Eq. \eqref{eqHRingApbc}. This term generates a global $\sigma_z$ coupling which does not commute with the Ising \textit{n.n.} interaction. At high transverse field value, i.e. $\delta/\tilde{\Omega}\gg1$ the initial state, $\ket{0000}$ becomes an eigenstate. Thus, approaching this limit, we expect an effective slowing down of the observed dynamics. 

\begin{figure}[H]
    \centering
    \includegraphics[width=0.81\linewidth]{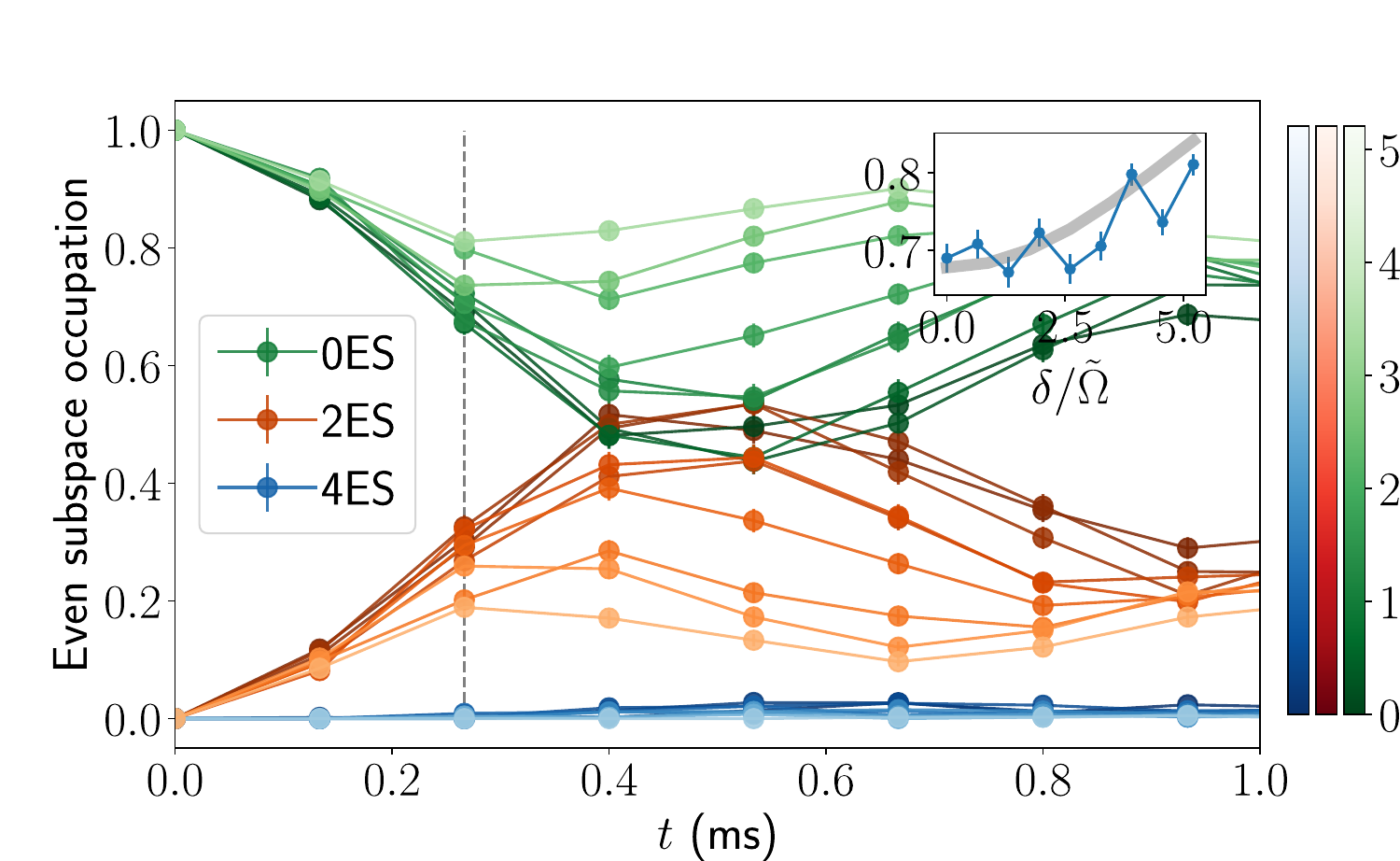}
    \caption{Dynamics of an anti-periodic Ising-spin ring, with a transverse field. We evolve the state $\ket{0000}$ under the system's Hamiltonian. Populations (points) and grouped to their ES (legend), and post-selected to even subspaces. We repeat this measurement for various magnitudes of transverse fields, $\delta/\tilde{\Omega}$ (color bars). For $\delta\gg\tilde{\Omega}$ the initial state becomes an eigenstate, leading to a suppression of dynamics. Error bars reflect $2\sigma$ errors due to quantum projection-noise. Inset: Measured ground state population (blue) after a short evolution time, $t=4\pi\xi^{-1}$ (vertical-grey), compared to an exact simulation (grey), showing good agreement.}
   \label{figTransverse}
\end{figure}

Figure \ref{figTransverse}, shows the post-selected populations in even excitation subspaces (points connected by lines), for various choices of $\delta/\tilde{\Omega}$ (color bars). Indeed, the ground state (green lines) remains more populated throughout the observed dynamics for larger values (brighter) of $\delta/\tilde{\Omega}$, while two excitations (orange) and four excitations (blue) become less frequent.

To analyze the data further, and benchmark our simulation, we consider the population in the ground state, $\text{Pr}\leri{0000}$, after two consecutive applications of $U$ in Eq. \eqref{eqUMulti} and the transverse field term (dashed gray line). Since $\delta$ acts to detune the \textit{n.n.} interaction, similar to off-resonance Rabi oscillations, we expect a quadratic suppression of dynamics. Indeed, the inset of Fig. \ref{figTransverse} shows the measured ground state population (blue), which exhibits a quadratic increase as a function of $\delta$. We also plot the theoretical prediction by directly simulating the Hamiltonian in Eq. \eqref{eqHRingApbc} (grey) showing a qualitative correspondence. We remark that for large values of $\delta$ the Trotter approximation becomes less-accurate which may account for the discrepancy between data and theory at $\delta/\tilde{\Omega}\approx 5$.

In conclusion, we have demonstrated a method for programmable quantum simulations of spin-Hamiltonians on trapped-ion chains. The method, based on the proposal in Ref. \cite{shapira2020theory}, can be used to generate a variety of models and coupling geometries, which are unconstrained by the physical realization of the 1D linear ion-crystal. Here we employ it to generate a 4-site anti-periodic Ising-spin ring. We use population and correlation data in order to benchmark our simulation. Indeed we reconstruct the applied couplings and transverse field, showing a faithful generation of the intended model. This method is well-suited for NISQ-era quantum systems as it can be used to leverage typical shallow circuits yet still generate long evolution times of relevant quantum systems.

During the preparation of this manuscript we became aware of a similar works by Wu et al. \cite{wu2023qubits} and Lu et al. \cite{lu2023realization}, which use similar techniques in order to implement similar Ising models.

\begin{acknowledgments}
This work was supported by the Israeli Science Foundation, the Israeli Ministry of Science Technology and Space, the Minerva Stiftung, the European Union’s Horizon 2020 research and innovation programme (Grant Agreement LEGOTOP No. 788715), the DFG (CRC/Transregio 183, EI 519/7-1), ISF Quantum Science and Technology (2074/19).
\end{acknowledgments}

\bibliography{main.bib}

\end{document}


\title{Programmable quantum simulations on a trapped-ions quantum simulator with a global drive - Supplemental Material}

\author{Yotam Shapira$^{1,\ast}$}
\author{Jovan Markov$^{1,\ast}$}
\author{Nitzan Akerman$^{1}$}
\author{Ady Stern$^{2}$}
\author{Roee Ozeri$^{1}$}

\affiliation{\small{$^1$Department of Physics of Complex Systems, Weizmann Institute of Science, Rehovot 7610001, Israel\\
$^2$ Department of Physics of Condensed Matter Systems, Weizmann Institute of Science, Rehovot 7610001, Israel\\
$^*$ These authors contributed equally to this work
}}

\maketitle

\onecolumngrid
\section{Constant phase shift freedom}
We derive the property used in main text in order to decouple from the center-of-mass mode of motion, i.e. we show that in Eq. (2) of the main text a shift of the form $\Phi_j\mapsto\Phi_j+\varphi$ for all $j=1,...,N$ has no physical effect on the simulated model.

We do so directly, starting from Eq. (2) of the main text,
\begin{align}
    U&=\exp\leri{i\sum_{j=1}^N\left(\Phi_j+\varphi\right)\sum_{n,m=1}^N O_j^{\leri{n}}O_j^{\leri{m}}\sigma_x^{\leri{n}}\sigma_x^{\leri{m}}}\\
    &=\exp\leri{i\sum_{j=1}^N\Phi_j\sum_{n,m=1}^N O_j^{\leri{n}}O_j^{\leri{m}}\sigma_x^{\leri{n}}\sigma_x^{\leri{m}}}\exp\leri{i\varphi\sum_{n,m=1}^N \sigma_x^{\leri{n}}\sigma_x^{\leri{m}}\sum_{j=1}^N O_j^{\leri{n}}O_j^{\leri{m}}}\\
    &=\exp\leri{i\sum_{j=1}^N\Phi_j\sum_{n,m=1}^N O_j^{\leri{n}}O_j^{\leri{m}}\sigma_x^{\leri{n}}\sigma_x^{\leri{m}}}\exp\leri{i\varphi\sum_{n,m=1}^N\delta_{n,m}\sigma_x^{\leri{n}}\sigma_x^{\leri{m}}}\\
    &=\exp\leri{i\sum_{j=1}^N\Phi_j\sum_{n,m=1}^N O_j^{\leri{n}}O_j^{\leri{m}}\sigma_x^{\leri{n}}\sigma_x^{\leri{m}}}\exp\leri{i\varphi\sum_{n=1}^N 1}\\
    &=\exp\leri{i\sum_{j=1}^N\Phi_j\sum_{n,m=1}^N O_j^{\leri{n}}O_j^{\leri{m}}\sigma_x^{\leri{n}}\sigma_x^{\leri{m}}}\exp\leri{i N \varphi},\label{eqSupDerivation}
\end{align}
where in the second line we split the exponential to a product of two exponentials as the Pauli-$x$ operators commute, in the third line we used the orthonormality of $O$, i.e. $\sum_{j=1}^N O_j^{\leri{n}} O_j^{\leri{m}} =\delta_{n,m}$, with $\delta_{n,m}$ the Dirac $\delta$-function, in the fourth line we evaluate the $\delta$-function and used the fact that $\leri{\sigma_x}^2=1$ and in the fifth line we evaluate the trivial summation on $n$.
This derivation shows that a constant shift of all modes by $\varphi$ amounts to a global phase shift of $U$ by $N\varphi$, which physically has no effect on dynamics. 

So far we have repeated the arguments given in Ref. \cite{shapira2020theory}, in which this degree of freedom has been utilized in order to perform an efficient optimization and norm-reduction of the resulting drive. 

We note that the shift freedom was enabled by the orthogonality of the normal modes of motion, given by $O$, and not by their normalization, i.e. the crucial component for this freedom is the cancellation of $n\neq m$ terms in the summation, due to the $\delta$-function. We note that any transformation of the form, $O_j^{\leri{n}}\mapsto\tilde{O}_j^{\leri{n}}=O_j^{\leri{n}}g\leri{n}$, with $g\leri{n}$ an arbitrary ion-dependent (and mode-independent) function will continue to yield this result. That is, for $n\neq m$ we have,
\begin{equation}
    \sum_{j=1}^N \tilde{O}_j^{\leri{n}} \tilde{O}
_j^{\leri{m}} =g\leri{n}g\leri{m}\sum_{j=1}^N O_j^{\leri{n}} O_j^{\leri{m}} =g\leri{n}g\leri{m}\delta_{n,m}=0\label{eqSupOrtho}.
\end{equation}

This generalization is helpful as it enables to correct for beam aberrations, described in the next section, while still utilizing the constant phase shift freedom in order to, e.g. decouple from the center-of-mass mode. 

\section{Beam aberration correction and entanglement phases design}
As we state in the main text, an inhomogeneity of the global driving field can be mitigated by our method. Indeed this can be done by effectively modifying the orthonormal mode matrix, $O$, as described in the section above. 

Specifically, we assume that the Rabi frequency associated with the $n$th ion is given by, $\Omega_n=\Omega_0 g\leri{n}$, with $\Omega_0$ a characteristic Rabi frequency and $g\leri{n}$ an ion-dependent, dimensionless, function that describes, e.g. a Gaussian envelope of the driving field. We define, $\tilde{O}_j^{\leri{n}}=O_j^{\leri{n}}g\leri{n}$ as a modified mode-matrix and set the phase accumulated by each mode according to $\tilde{O}$ (instead of $O$ as usual). As described in the section above, this modification of $O$ still allows to decouple from the center of mass mode of motion. However the mode-dependent phases, $\Phi_j$, are now designed in accordance with this modification, thus the beam's profile is optimally mitigated, within the scope of the homogeneous drive.

The choice of the $\Phi_j$s continues as prescribed in Ref. \cite{shapira2018robust}. Specifically we formulate the set of equations,
\begin{equation}
    \sum_j\Phi_j \tilde{O}_j^{\leri{n}}\tilde{O}_j^{\leri{m}}=J_{n,m},
    \label{eqSupLinear}
\end{equation}
for all $n,m=1,...,N$, with $J_{n,m}$ some desired coupling matrix. This constitutes an over-determined linear system for $\boldsymbol{C}=L \boldsymbol{\Phi}$, with $\boldsymbol{\Phi}=\left\{\Phi_j\right\}_{j=1}^N$ and $\boldsymbol{C}$  and $L$ read directly from Eq. \eqref{eqSupLinear}. 

This linear equation can be solved by pseudo-inverting $L$, or by any other standard technique, yielding $\boldsymbol{\Phi}$ of the main text. We proceed to choose tone frequencies and amplitudes such that each mode is coupled by 4 tones: $\omega_0\pm\leri{\nu+\xi}$ and $\omega_0\pm\leri{\nu+3\xi}$ with the amplitude of the latter pair equal in amplitude and phase-shifted by $\pi$. The amplitude of the tones associated with the $j$th mode are proportional to $\sqrt{\Phi^\text{optimal}_j}$ with the proportionality factor $\tilde{\Omega}$, described below. With these choices the system's evolution agrees with the underlying spin-model at stroboscopic times, integers of $2\pi\xi^{-1}$. Here we have chosen to operate in the adiabatic regime, i.e. the stroboscopic time is considerably longer the the period of oscillation of all modes of motion, $\xi\ll\nu_j\;\forall j$. In our system we have $\nu=\leri{0.819,1.419,1.974,2.499}$ MHz, measured independently, with estimation error $<100$ Hz.

It is possible to increase the simulations speed by using a higher Rabi frequency, however in the fast regime the relation between $\boldsymbol{\Phi}^\text{optimal}$ and the tones structure is no-longer trivial. Obtaining it involves solving a non-trivial optimization problem. Nevertheless, a method to generate such solutions appear in Ref. \cite{shapira2023fast} which showcases solutions of coupling maps for ion-crystals of up to $N=100$ ions.

\section{Dynamical simulation and fitting}
Figure 1 of the main text presents dynamics data accompanied by a theoretical prediction and a simulation, we describe details of the simulation.

In our system, the qubit-qubit interactions are mediated the global modes of motion of the ion-crystal. In principle our observed data can be generated by numerically evolving such a system, however the size of Hilbert space, already for $N=4$ ions, is large and makes such simulations challenging. Specifically, assuming each phonon mode is modeled using $N_c=8$ phonon levels, then the Hilbert state is of dimension $\leri{2N_c}^N$, i.e. evolving 65536 complex amplitudes for $N=4$. Since we would like to include also incoherent effect the resulting density matrix has more than $10^9$ entries. Furthermore, our aim is to perform a fit over the simulation parameters, making this process infeasible.

Instead, we take a similar approach used in Ref. \cite{shapira2023quantum}, and describe only the qubit evolution in relevant subspaces. Effects due to the phonon degrees are freedom are modeled as Lindbladian noise channels.

We consider the evolution of a density matrix, $\rho$, in the Lindblad equation, 
\begin{equation}
    \partial_t{\rho}=-i\left[H,\rho\right]+\sum_i\gamma_i\leri{L_i\rho L_i^\dagger-\frac{1}{2}\left\{L_i^\dagger L_i,\rho\right\}}, \label{eqSupLindblad}
\end{equation}
with $H$ a system Hamiltonian similar to $H_\text{ring,a.p}$ in Eq. (4), i.e. $H=\sum_{n,m}^4J_{n,m}\sigma_x^{\leri{n}}\sigma_x^{\leri{m}}+\delta\sum_{n=1}^4\sigma_z^{\leri{n}}$. The $L_i$s are Lindblad jump operators acting with rates $\gamma_i$.

As also shown in Ref. \cite{shapira2023quantum}, the effects of spin and motional dephasing on the qubit-qubit interaction is a slow and incoherent increase of the qubit population in the odd excitation subspaces. Since interactions between even-occupation states are mediated through the odd-subspace, their occupation is translated directly to decoherence in the even-subspace states. 

The population of the odd subspaces is modeled as an incoherent increase, at short evolution times this yields, $P_\text{odd}\leri{t}=1-e^{-\gamma_\text{odd} t}$ with $\gamma_\text{odd}$ determined by the incoherent sum of couplings of the drive to odd-excitation states,

\begin{equation}
    \gamma_\text{odd}=N T_2^{-1}\Omega^2 \sum_n \frac{\eta_n^2r_n^2}{\xi_n^2}
\end{equation}
with $T_2$ the single-qubit dephasing time, $\Omega$ a characteristic single-qubit Rabi frequency, $\eta_n$ the Lamb-Dicke parameter of the mode coupled by the $n$th tone, which is detuned from it by $\xi_n$ and has relative amplitude $r_n$, with $\sum_n r_n=1$.

With the parameters used in our experiment, and $T_2\approx5$ ms, measured independently, we obtain the estimate, $\gamma_\text{odd}\approx56$ Hz

The odd subspace populations generates dephasing between states that are coupled by the qubit-qubit interaction. This is modeled by jump operators of the form, $L_{n_\text{e}}=|n_\text{e}\rangle\langle n_\text{e}|$ and rate $\gamma_{n_\text{e}}\leri{t}=P_\text{odd}\leri{t}T_2^{-1}$, with $|n_\text{e}\rangle$ a $N$-qubit state with an even number of excitations. In total 8 such operators are used.

In addition to this effective `phonon-mediated' dephasing, direct dephasing, due to $T_2$, between different excitation subspaces occurs, which is modeled by $L_s=2sP_s$, and rate $T_2^{-1}$, with $P_s$ a projection operator on the $s$ES, with $s=\leri{0,2,4}$.

We obtain the population in the various $2^N$ basis states by numerically evolving $\rho$. We perform a fit by minimizing the difference between these simulated populations and the measured data. The fit is performed over the 6 values of $J_{n,m}$ as well as $T_2$. The resulting fit yields a mean population error per state per data point of $\langle\epsilon\rangle=0.0154(7)$. We remark that without inclusion of dephasing, i.e. by forcing $T_2^{-1}=0$, the fitting error doubles.

The resulting population, grouped to excitation subspaces are shown in Fig. 1 
 of the main text (solid). In our simulation odd excitation subspaces remain decoupled, and are therefore not fitted, instead we plot $P_\text{odd}$ with the fitted value of $T_2$ in the same figure (brown and olive) showing a good fit to the data.

The fit yields, $T_2^{-1}=204\pm8$ Hz, and the values of $J_{n,m}$ (in kHz) are $J_{1,2}=1.83$, $J_{1,3}=-0.04$, $J_{1,4}=-1.69$, $J_{2,3}=1.37$, $J_{2,4}=0.00$, $J_{3,4}=1.85$, with a maximum error of 11 Hz on all of these estimates. Here error estimation is based on a combination of fitting error and measurement error of the fitted data. These results yield a characteristic Rabi frequency of $\tilde{\Omega}\equiv\sum_{n>m}|J_{n,m}|/6\approx1.13$ kHz. We use the same fidelity test for the fitted $J_{n,m}$ values and obtain, $F\leri{J_{n,m}^{\text{optimal}},J_{n,m}^\text{fit}}=0.994\substack{+0.006 \\ -0.01}$ and $F\leri{J_{n,m}^{\text{reconstructed}},J_{n,m}^\text{fit}}=0.986\substack{+0.014 \\ -0.042}$, in agreement with the independent parity measurements.

\section{Correlation function computation}
We show that the parity measurement, described by the unitary evolution, 
\begin{equation}
    U_\text{cor}=e^{\frac{i}{2}\frac{\pi}{2}\sum_{n=1}^N\sigma_\phi^{\leri{n}}} e^{i\sum_{n,m}J_{n,m}X_n X_m}\label{eqSupUParity},
\end{equation}
yields measurement of $J_{n,m}$, in leading order in the $J_{n,m}$s. Note that compared to Eq. (5) of the main text absorbed a factor of $\Omega t_\text{cor}$ in the $J_{n,m}$.

We have,
\begin{equation}
    C_{n,m} =\langle\psi|Z_n Z_m|\psi\rangle=\langle0000|U_\text{cor}^\dagger Z_n Z_m U_\text{cor}|0000\rangle 
\end{equation}
Next, we note that,
\begin{equation}
    U_\text{cor}^\dagger Z_n Z_m U_\text{cor}=e^{-i\sum_{n,m}J_{n,m}X_n X_m}\sigma_{\phi^\prime}^{\leri{n}}\sigma_{\phi^\prime}^{\leri{m}}e^{i\sum_{n,m}J_{n,m}X_n X_m},
\end{equation}
with $\phi^\prime=\phi-\pi/2$ and ,
\begin{align}
    \sigma_{\phi^\prime}^{\leri{n}}\sigma_{\phi^\prime}^{\leri{m}}&=\cos^2\leri{\phi^\prime} X_n X_m+\sin^2\leri{\phi^\prime} Y_n Y_m \\
    &+\cos\leri{\phi^\prime}\sin\leri{\phi^\prime}\leri{X_n Y_m+X_m Y_n}
\end{align}
Assuming the $J_{n,m}$s are small we expand,
\begin{equation}
    e^{i\sum_{n,m}J_{n,m}X_n X_m}=1+i\sum_{n,m}J_{n,m}X_n X_m+\mathcal{O}\leri{J_{s,t}^2},
\end{equation}
with $J_{s,t}$ meaning all coupling terms in $J$. With this expansions only the $Y_n Y_m$ evaluate to non-vanishing terms. We obtain,
\begin{equation}
    C_{n,m}=2J_{n,m}\sin\leri{2\phi^\prime}+\mathcal{O}\leri{J_{s,t}^3},
\end{equation}
which justifies our parity fringe fit in the main text.
As mentioned in the main text, inhomogeneity in the driving field may cause a bias in $C_{n,m}$ due to errors in the analysis pulse. We model this by generalizing $U_\text{cor}$ to,
\begin{equation}
    U_\text{cor}=e^{\frac{i}{2}\frac{\pi}{2}\sum_{n=1}^N\leri{1+\varepsilon_n}\sigma_\phi^{\leri{n}}} e^{i\sum_{n,m}J_{n,m}X_n X_m}\label{eqSupUParity},
\end{equation}
with $\varepsilon_n$ the relative inhomogeneity on the $n$th ion. Repeating the analysis above this yields,
\begin{equation}
    C_{n,m}=\leri{2J_{n,m}-\frac{\pi^2}{4}\leri{\varepsilon_n^2+\varepsilon_m^2}}\sin\leri{2\phi^\prime}+\frac{\pi^2}{4}\varepsilon_n\varepsilon_m,
\end{equation}
with corrections that are cubic in $J_{s,t}$, $\varepsilon_n$ and $\varepsilon_m$.
\bibliography{main.bib}